\documentclass[prl,twocolumn,showpacs,preprintnumbers,amsmath,amssymb,superscriptaddress]{revtex4}
\usepackage{graphicx}
\usepackage{dcolumn}
\usepackage{bm}
\usepackage{psfrag}

\begin{document}
\title{Incommensurability and spin excitations of diagonal stripes in cuprates} 
\author{G. Seibold}
\affiliation{Institut f\"ur Physik, BTU Cottbus, PBox 101344,
03013 Cottbus, Germany}
\author{J. Lorenzana}
\affiliation{Center for Statistical
Mechanics and Complexity, INFM, Dipartimento di Fisica,
Universit\`a di Roma La Sapienza, P. Aldo Moro 2, 00185 Roma, Italy}
\date{\today}
\begin{abstract}
Based on the time-dependent Gutzwiller approximation
we study the possibility that the diagonal incommensurate spin scattering
in strongly underdoped lanthanum cuprates originates from
antiferromagnetic domain walls (stripes). 
Calculation of the dynamic spin response for stripes in the diagonal
phase yields the characteristic hour glass dispersion with the crossing
of low energy Goldstone and high-energy branches at a characteristic
energy $E_{cross}$ at the antiferromagnetic wave-vector $Q_{AF}$. 
The high energy part is close to the parent antiferromagnet. 
Our results suggest that inelastic neutron scattering experiments 
on strongly underdoped lanthanum cuprates can be understood as due to a 
mixture of 
bond centered and site centered stripe configurations with 
substantial disorder. 
\end{abstract}


\maketitle

The low energy spin response of many high-T$_c$ superconductors 
is characterized by magnetic fluctuations which are incommensurate
with respect to the antiferromagnetic (AF) order 
(for a review see \cite{tran05}). 
Since in  codoped (with Eu, Nd or Ba) lanthanum cuprates (LSCO) 
an accompanying charge-density wave has been detected (e.g. \cite{abba05})
this kind of scattering is usually attributed
to the formation of stripe textures where the doped holes
segregate into quasi one-dimensional 'rivers of charge' which simultaneously
constitute domain walls for the AF order parameter.

The similarity of low-energy inelastic neutron scattering (INS) data between 
codoped  and non-codoped materials supports  the picture of
a fluctuating and (or) glassy stripe phase in LSCO. 
From these experiments it turns out that above doping
$n_h=0.055$ the stripes are
oriented along the Cu-O bond direction ('vertical stripes') 
and the incommensurability $\delta$
(defined as the deviation of the 
magnetic peak from $Q_{AF}$) linearly increases up 
to $n_h\approx 1/8$~\cite{yam98}.
Above $n_h\approx 1/8$, $\delta$ stays essentially constant but the intensity
of the low energy spin fluctuations decreases and vanishes at the same
concentration where superconductivity disappears in the overdoped regime
thus suggesting an intimate relation between both phenomena \cite{waki04}.

Upon lowering the doping below $n_h\approx
0.055$ the incommensurability $\delta$
rotates by $45^0$ \cite{waki99,waki00,mats00,fujita02}
to the
diagonal direction and the orthorhombic lattice distortion 
allows one to conclude that
the elastic diagonal magnetic scattering is one-dimensional with the associated
modulation along the orthorombic $b^*$-axis.
When $\delta$ is measured in units of  reciprocal tetragonal
lattice units in both the vertical and diagonal phase it turns out that 
the magnitude of the incommensurability numerically coincides across
the rotation leading to a linear relation  $\delta=x$.
Upon approaching the border of the AF phase at $n_h=0.02$ 
the incommensurability 
approaches $\epsilon=x$ \cite{matsuda02} where $\epsilon$ is 
measured in units of  
reciprocal orthorombic lattice units, thus $\epsilon=\sqrt{2}\delta$.

In this paper we present computations of the incommensurability and
the dynamic response of diagonal stripes based on the Gutzwiller
approximation and its dynamical extension and compare with our previous
computations for vertical stripes and recent INS data \cite{matsuda08}. 
Our results suggest that in the
spin glass phase incommensurate scattering is due to a disordered
mixture of bond centered and site centered diagonal stripe configurations. 
\begin{figure}[tbp]
\includegraphics[width=8cm,clip=true]{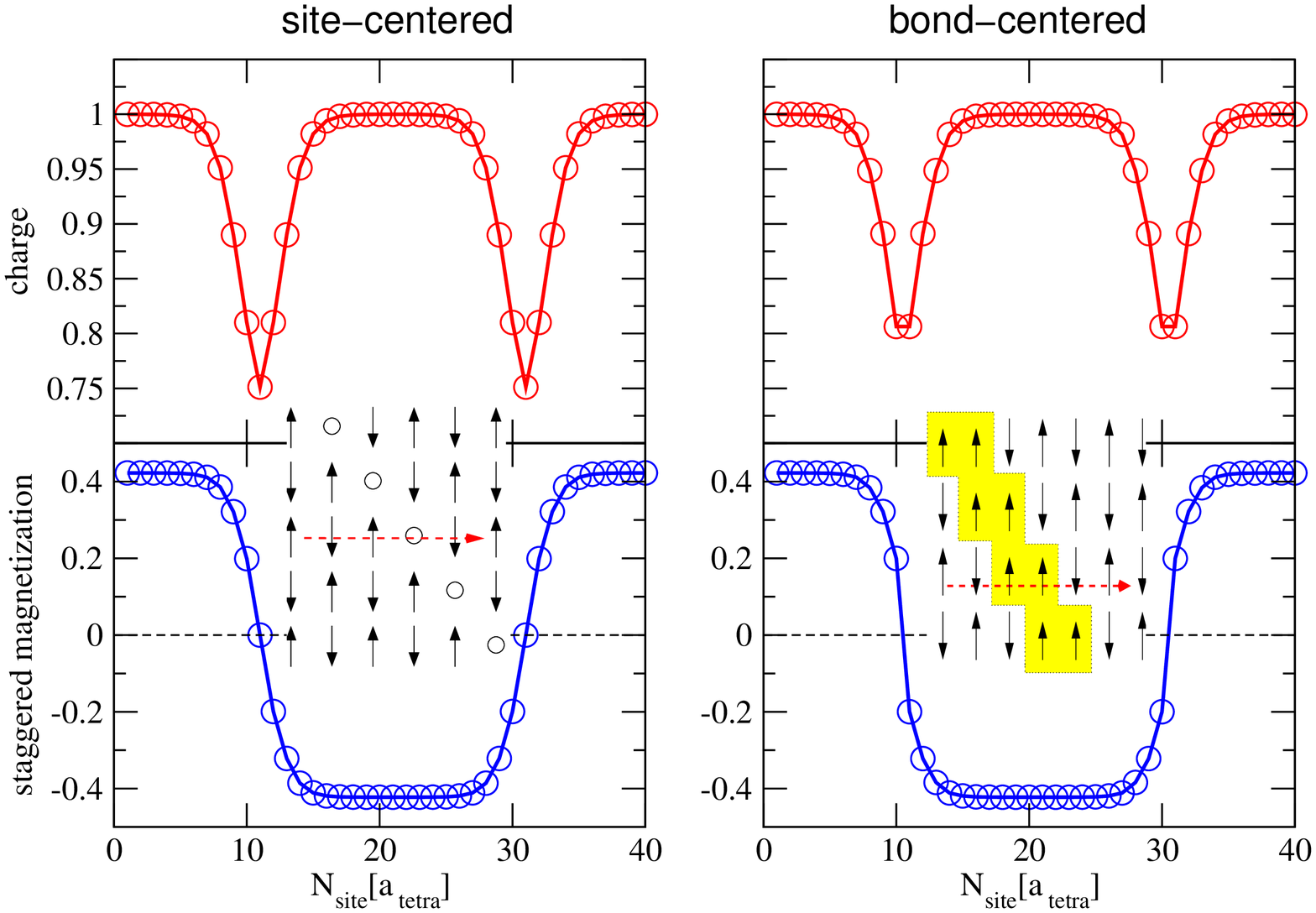}
\caption{(Color online) Charge density and magnetization for site centered and
bond centered diagonal stripes. The densities are plotted along
an horizontal scan, indicated by the dashed arrow.}
\label{fig1}
\end{figure}
Calculations are based on the one-band Hubbard model 
(on-site repulsion $U$) with
hopping restricted to nearest ($\sim t$) and next nearest ($\sim t'$)
neighbors. 
We first treat the model within an unrestricted Gutzwiller
approximation (GA) which leads to the stripe textures\cite{sei98}.
On top of this we obtain the spin excitations within the time
dependent Gutzwiller approximation\cite{sei01,sei04a}.
Parameters are $U/t=7.5$, $t'/t=-0.2$ and 
$t = 340 meV$\cite{sei06}. In previous 
papers we have shown that our approach can reproduce both, 
the magnon excitations of undoped LCO\cite{sei06} as revealed by
neutron scattering\cite{col01} and
the doping dependent incommensurability in the vertical stripe
phase\cite{lor02b,sei04}. In addition, the formalism leads to quantitative 
agreement  with the dispersion of 
spin excitations in LBCO\cite{sei05,sei06,lor05} and can reproduce the 
doping dependence of the optical conductivity\cite{lor03}. 

Fig.~\ref{fig1} shows the charge and magnetization profile
for the diagonal stripe structures we have obtained
as stable saddle-point solutions within the GA. The first is a site centered
diagonal (DSC) stripe where the doped holes are confined to
nonmagnetic sites which simultaneously constitute the antiphase
boundary for the AF order parameter. 

The second stable texture (Fig. \ref{fig1}, right panel) can be considered 
as the diagonal counterpart to vertical bond-centered (VBC) stripes. 
However, due to the diagonal orientation the stripe acquires a 
net ferromagnetic moment. Moreover, the magnetization points in the
same direction (i.e. the whole layer becomes ferromagnetic) when the BC
stripes are separated by an odd number of lattice constants in x- (or y-)
direction. They can be considered as the smallest 
'staircase' variant as discussed by Granath \cite{granath04}. We have
found that more extended 'staircase' structures are higher in energy and 
do not yield stable magnetic excitations.

The approximate linear relation between incommensurability and filling
at low doping\cite{yam98,waki99,waki00,mats00,fujita02} implies that
isolated stripes are self-bounded linear aggregations of holes with a
well defined number of holes per Cu along the stripe given by 
$\nu_{opt}\equiv n_h/\epsilon$. Indeed as for vertical
stripes\cite{sei04,lor02b} we find an optimum filling, however, an
important difference here is that $\nu_{opt}$ depends substantially
on the texture.  DSC are insulating (the Fermi level falls in a gap)
with $\nu_{opt} \approx 1$, thus 
$\epsilon \approx n_h$. Diagonal BC (DBC) stripes are metallic and
have an optimum filling at 
$\nu_{opt} \approx 0.75$ implying $\delta \approx n_h$  \cite{erice08}.

While DBC stripes are practically (accidentally) degenerate in energy
with low doping VBC, the energy of the DSC texture is 
$\approx 0.02t$ per hole above \cite{erice08}. We believe that these small
energy differences are not significant given the simplicity of the model.
A precise determination of the relative
stability of the different phases would require at least multiorbital
effects, inclusion of both long-range Coulomb interactions
and coupling of the holes to the tilts of the CuO$_4$ octahedra. The latter
have been shown to play a major role in the stabilization of
vertical {\it vs.} diagonal stripes\cite{normand01}. On the other
hand we believe that results within one phase are much more reliable. 
For example  
we have checked that the different optimum $\nu$ for DSC and BSC
stripes for a 3-band hamiltonian do not differ from those found for
the one-band model so that we consider this as a robust feature
of our calculation. 

In the present case the more important extrinsic effect is
disorder.  For vertical stripes the
correlation length can reach $150 a_{tetra}$ or around 20 times the
magnetic stripe periodicity while for diagonal stripes the correlation
is of the same order or even smaller than the
periodicity\cite{waki00}. This should be kept in mind while comparing
with our results which correspond to perfectly ordered stripes arrays. 
\begin{figure}[t]
\includegraphics[width=8cm,clip=true]{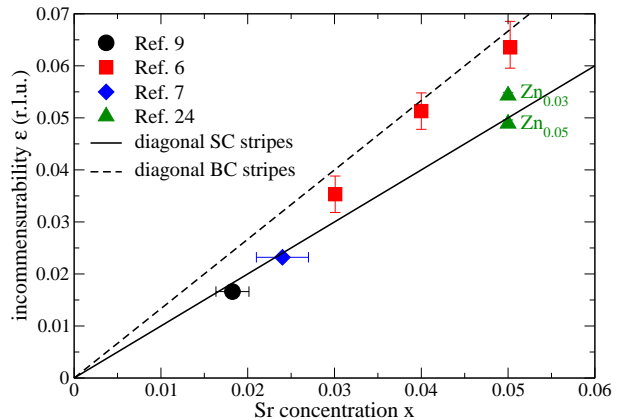}
\caption{(Color online) 
Doping dependence of the magnetic incommensurability $\epsilon$
in the diagonal phase as deduced from the referenced elastic neutron scattering
experiments. Solid and dashed line are the predicted dependence for 
DSC and DBC stripes, respectively.}
\label{fig2}
\end{figure}
Fig.~\ref{fig2} compares the doping dependence of the 
magnetic incommensurability $\epsilon$ in the diagonal phase as seen 
in elastic neutron scattering experiments \cite{waki00,mats00,matsuda02,matsuda06} together with the predicted
dependence  for DSC and DBC stripes. For non-codoped samples the
experimental incommensurability is in between the theoretical ones for
DBC and DSC, with a stronger tendency for DBC close to the 
metal-insulator transition which shifts to DSC at low doping.  
An interesting effect occurs upon substitution of  Cu$^{2+}$ by
nonmagnetic Zn$^{2+}$ which has a filled $3d^{10}$
shell\cite{matsuda06}. 
This should favor DSC stripes which are forced to have zero magnetic
moments in the core and indeed the incommensurability shift to the
computed DSC line (c.f. Fig. \ref{fig2}) is in accord with our
expectation.  All this supggests that the system is made of a disordered
mixture of DSC and DBC configurations with a gradual shift from the
former to the latter as doping is increased. 
\begin{figure}[t]
\includegraphics[width=8.5cm,clip=true]{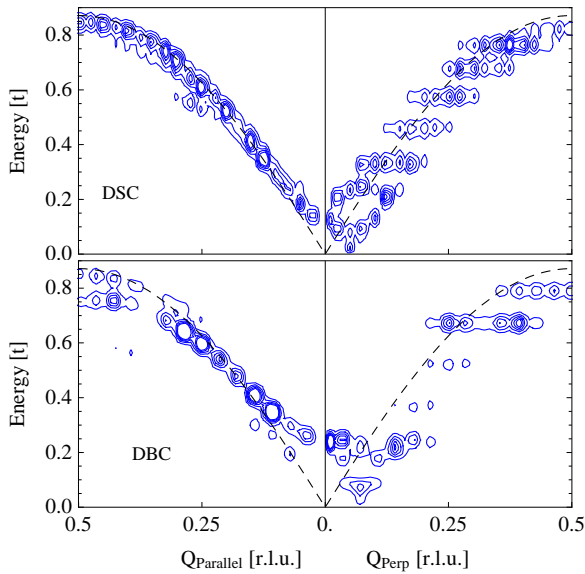}
\caption{(Color online)
Magnetic excitations perpendicular
and parallel to DSC (upper panel) and DBC (lower panel) 
stripes (doping $n_h=0.05$) ad reveled by a contour plot of  
$I(\omega,{\bf q})\equiv \omega S(\omega,{\bf q})$. The momentum scale
is the distance from $Q_{AF}$ measured in units of $2\pi/a_{ortho}$.
}
\label{fig3}
\end{figure}

In the following we discuss the spin excitations for doping
$n_h=0.05$. For DSC stripes we have $\epsilon=0.05$ (i.e. charged stripes
separated by $20 a$ along the x-, or y-direction and for DBC stripes with
$\epsilon=0.071$ (i.e. separated by $14 a$ along x-, or y-direction).
Fig.~\ref{fig4} displays $I(\omega,{\bf q})\equiv \omega S(\omega,{\bf q})$
with  $S(\omega,{\bf q})$ the transverse dynamical structure factor 
relevant for magnetic neutron scattering experiments.  The $\omega$
factor makes visible the high energy features which would appear much
faint in the experimentally accessible $S(\omega,{\bf q})$.
 We show the 
dispersions of magnetic excitations perpendicular (left panel)
and parallel (right panel) to the stripe direction.

Parallel to the stripe the dispersion is dominated by modes very
similar to the ones of the undoped AF as could be expected from the real
space structure.  
Due to the large unit cell and the small extension of the reduced
Brillouin zone in the perpendicular direction excitations consist of a
large number of folded
bands. These are generally silent when plotted in the extended
Brillouin zone but acquire a large spectral weight in proximity to the 
modes of the parent antiferromagnet (shown with a dashed line) giving
rise to the horizontal segments in the right panel. This is analogous
to the effect reported in photoemission where spectral weight is maximized when
intersecting the free electron dispersion
relation\cite{granath04}. This clarifies the evolution of the
spectrum at very low doping: As doping
decreases the dynamical structure factor gradually tends to the one of
the antiferromagnet by increasing the number of bands and modulating the
spectral weight.  

At low energies the
spectrum is determined by the Goldstone excitations belonging to an
inwards (i.e. in the direction of $[\pi,\pi]$) and an outwards dispersing
branch. 
For DSC stripes upon increasing energy the
inwards dispersing branch dominates in weight and reaches the AF
wave-vector $Q_{AF}$ at $E_{cross}\approx 35 meV$ where it connects 
to the dispersion along the stripe. 
In contrast to the vertical excitations \cite{sei06}
the saddle-point structure is very small and barely resolved 
due to finite size effects. For DBC stripes a small gap appears
between the Goldstone mode and the saddle point in strong resemblance 
with linear spin-wave theory (LSWT) results by
Carlson {\it et al.} 
\cite{carlson04}.  There it was pointed out, that for even spaced
diagonal stripes (as in the present case) $Q_{AF}$ becomes a
reciprocal lattice vector with a downturn of the acoustic branch. Interestingly
for DSC stripes in our itinerant approach the analogous gap is not
resolved resulting in practice in a continuous dispersion of the
Goldstone mode up to the saddle with energy $E_{cross}$. We expect
that in both cases the gap will be washed out by disorder and damping
effects. 
\begin{figure}[t]
\includegraphics[width=8.5cm,clip=true]{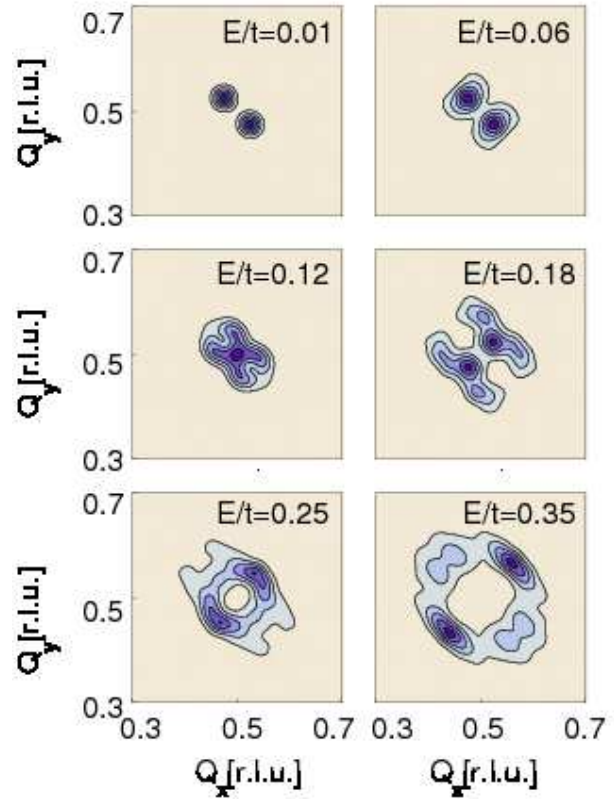}
\caption{(Color online) Constant frequency scans of $Im \chi_q{\omega}$ 
for DSC stripes ($n_h=0.05$). Data have been convoluted by a gaussian in
frequency and momentum space which half-width corresponds to the resolution
of the calculation, i.e. $\delta\omega=0.01 t$ and $\delta q = 2\pi/80 1/a_{tetra}$. The center of each panel corresponds to $Q_{AF}$ and wave-vectors
are denoted in units of $2\pi/a_{tetra}$. 
}
\label{figscan}
\end{figure}

In Fig.~\ref{figscan} we show cuts of the dispersion for the DSC
stripes. At low frequencies the excitations at the two incommensurate
wave-vectors carry  
the dominant weight which start to form spin-wave cones upon increasing energy.
The cones merge at $E_{cross}=0.12t$ where it is apparent that the weight of the
cones is strongly anisotropic and enhanced close to $Q_{AF}$. Similar to our
previous findings for vertical stripes \cite{sei05,sei06} the response well
above $E_{cross}$ becomes two-dimensional forming a ring shaped pattern 
around $Q_{AF}$.
\begin{figure}[htbp]
\includegraphics[width=8cm,clip=true]{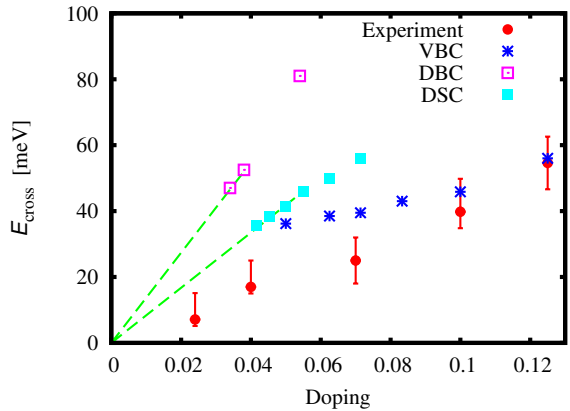}
\caption{(Color online) Doping dependence of $E_{cross}$ for DSC  and DBC
 stripes. Stars refer to results for vertical
stripes and squares to diagonal ones. Circles with error bars are
experimental data from  Refs. \onlinecite{matsuda08,kofu07,tran04,hiraka01}.
 Dashed lines are the extrapolation of $E_{cross}$ to $\omega=0$ at $n_h=0$. 
}
\label{fig4}
\end{figure}

Fig.~\ref{fig4} reports the doping dependence of E$_{cross}$ for both
SC (main panel) and BC (lower right panel) textures in the vertical
and diagonal phase compared to INS data 
\cite{matsuda08,kofu07,tran04,hiraka01}.  Since for the undoped system 
one recovers the spin excitations  of the AF, $E_{cross}$ 
extrapolates down to zero frequency at $n_h=0$ (dotted line)
for both DSC (main panel) and DBC (lower right panel) stripes.
Due to the large supercells involved we could not access the doping
regime $n_h<0.04$. 

Below $x=0.08$ our computations overestimate E$_{cross}$. One should
keep in mind that we are working with a fixed set of parameters
determined in Ref.~\onlinecite{sei06}. For example reduction of  $t$ improves
considerable the agreement but spoils the fitting of the undoped
dispersion relation. 
Due to the complexity of the computations
we have not performed intense optimization of the parameter set, so it may
be that for different choices of $U$ and $t'$ and $t$ we could also
achieve a better fit of the overall doping dependence. However, we
believe the main reason for the disagreement is the strong
disorder character of the stripes in this doping range. Indeed quite
generally one expects a 
softening of excitations as the magnetic configuration becomes more
disordered. This becomes evident in the Ising limit comparing the
dispersion relation of an Ising ferromagnet (or AF) with that of an
Ising spin glass in which the exchange constant $J$ is the same but the
sign is random. In the ordered case the spin flip excitation energy is
$2 z J$ with $z$ the coordination, whereas in the glassy case it is
smaller in average due to frustrated configurations. 
We have checked for an array of stripes on a finite cluster that the
inclusion of random local impurity potentials leads to a softening
of E$_{cross}$, whereas a variation of the stripe spacing induces a
broadening of the corresponding excitation.

To conclude, we have shown that the doping dependence of the
incommensurability of diagonal magnetic scattering is compatible with
a mixture of SCD and BCD stripes. Assuming an ordered array of stripes 
we find excitations that are qualitatively in accord with experiment 
\cite{matsuda08}
but too high in energy. We attribute the difference to the much more
disordered nature of diagonal incommensurate scattering. 
Our intermediate coupling approach predicts a quasi two-dimensional
high energy response in the diagonal phase of LSCO which displays a
one-dimensional static response along the orthorombic $b^*$-axis. 
Our computations also clarify how the dispersion relation converges to the
one of the AF as doping is decreased. 

G.S. acknowledges financial support from the Vigoni foundation.

\end{document}